\documentclass[aps,prl,preprint,groupedaddress]{revtex4-2}
\usepackage[utf8]{inputenc}
\usepackage{physics}
\usepackage{mathrsfs}
\usepackage[colorlinks=true,citecolor=blue,linkcolor=blue]{hyperref}
\usepackage[english]{babel}
\usepackage[T1]{fontenc}
\usepackage[Lenny]{fncychap}
\usepackage[left=2cm,right=2cm,top=2cm,bottom=2cm]{geometry}
\usepackage{amsmath,amssymb,amsthm,dsfont}
\usepackage{amsfonts}
\usepackage{graphicx}
\usepackage{subfig}
\usepackage{wrapfig}
\usepackage{slashed}
\usepackage{fancyhdr}
\selectfont
\usepackage{babel}

\begin{document}


\title{Laser-Assisted Pion Decay}


\author{S. Mouslih$^{2,1}$, M. Jakha$^1$, S. Taj$^1$ and B. Manaut$^1$}
\email[]{b.manaut@usms.ma}
\affiliation{ $^1$ Universit\'e Sultan Moulay Slimane, Facult\'e Polydisciplinaire,\'Equipe de Recherche en Physique Th\'eorique et Mat\'eriaux (ERPTM), B\'eni Mellal, 23000, Morocco.}
\author{E. Siher$^2$}
\affiliation{$^2$ Faculté des Sciences et Techniques, 
		Laboratoire de Physique des Matériaux (LPM),
		Béni Mellal, 23000, Morocco.}

\date{\today}

\begin{abstract}
This paper revives the controversial debate that has arisen over the last two decades about the possibility that the electromagnetic field affects the lifetime or the decay rate of an unstable particle. In this research, we show, by performing analytical calculations and extracting numerical results, that the pion lifetime can be changed notably by inserting the decaying pion into an electromagnetic field only if the number of photons transferred between the decaying system and the laser field does not reach the well-known sum-rule. The influence of the laser parameters on the decay rate and branching ratio is also discussed. The surprising result obtained for the pion lifetime is referred to as the well-known quantum Zeno effect.
\end{abstract}

\maketitle
\section{Introduction}
The study of particle behavior and  their properties when they are inserted into an electromagnetic field has received much attention in recent times \cite{A,B}, due to the progress made by laser technology in terms of both its intensity and its sources \cite{C}. As one of these properties, we find the lifetime or the decay rate of a particle. In the meantime, the study of decay in the presence of an electromagnetic field and the size of the effect that the latter may have on the lifetime were and are still the subject of much scientific research \cite{Decays}. Thirteen years ago, in a controversial scientific research \cite{D}, Liu \textit{et al.} studied the effect of a strong laser field on the decay rate of the muon and found a dramatic change in its lifetime, as much as an order of magnitude. This result is criticized by Narozhny and Fedotov in their comment \cite{E,F} and they consider it to be invalid and contradictory to physical intuition. Two other authors have done their own calculation with different laser polarization and also reach a very different conclusion \cite{G,H}. They found the effect of the laser on the muon lifetime to be very small. In this paper, and in attempt to stimulate the debate again and deepen our understanding more about this controversy, we decided to study the decay of the charged pion, which is a composite particle completely different from the muon. The pion decay in a
linearly polarized laser field was thoroughly studied
almost $50$ years ago by Ritus in \cite{Ritus}. The weak decay processes in the presence of strong laser fields can be divided into two categories: first, laser-assisted processes which also exist in the absence of the field but may be modified due to its presence. Second, field-induced processes which can occur only when a background field is present, providing an additional energy reservoir \cite{Piazza}. The pion and muon decays belong to the first class. Apart from this context, the pions being the lightest mesons hold a special place in both the weak and the strong interactions, and remain subjects of research interest ever since their discovery almost $70$ years ago \cite{I}. Historically, pion decay has provided an important testing ground for the weak interaction and radiative corrections and was considered the best experimental evidence for the vector or axial-vector character of weak interactions \cite{J}. Pion decay also attracted a lot of interest from experimental research; for example in the article \cite{K}, the authors have presented an experimental study of rare charged pion decays. The muon momentum in the pion decay at rest has measured experimentally using a magnetic spectrometer by \cite{L}, while the precision measurements of the branching ratio $R_{e/\mu}$ between muon and electron decays have been presented in \cite{M,Brexp} and they provide the best test of $e$-$\mu$ universality in weak interactions. Our aim in this work is to study the effect of a circularly polarized laser field on the decay rate and the lifetime of the negatively charged pions, which are unstable and decay into two leptons $\pi^{-}\rightarrow e^{-}\bar{\nu}_{e}$ or $\mu^{-}\bar{\nu}_{\mu}$  by virtue of weak interaction with a lifetime $\tau_{\pi^{-}}=(2.6033\pm0.0005)\times 10^{-8}~\text{sec}$ \cite{PDG}. The laser-free pion decay can be found in many textbooks as \cite{Griffiths}. As mentioned previously in \cite{E}, there are two mechanisms of an external field influence on a particle decay. The first of them is purely kinematical and it is regulated by the parameter $\eta=e\mathcal{E}_{0}/m\omega$ (here, $m$ is the rest mass of the particle, $e$ is the electric charge, $\mathcal{E}_{0}$ is the electric field amplitude of the laser and $\omega$ its frequency), while the second  mechanism is governed by an independent dynamical parameter $\mathcal{X}=\mathcal{E}_{0}/\mathcal{E}_{S}$ where $\mathcal{E}_{S}=m^{2}c^{3}/e\hbar$. As long as these two parameters have very small values for the determined intensities, the decay rate of a free particle is unaffected by the presence of an external electromagnetic plane wave. In our case, the values of the two parameters mentioned above are very small since we use the laser field amplitude $10^{6}-10^{12}~\text{V/cm}$ and $m_{\pi}\simeq273~m_{e}$. Note that the laser is supposed to be turned on adiabatically for a period of time sufficiently longer than the laser-free pion lifetime, and the laser intensity used is taken so that it does not allow pair creation \cite{paircreation}. Therefore, we agree with Narozhny and Fedotov that there should not be an effect of the laser field on the pion lifetime. But, what we emphasize and what we will demonstrate from the results obtained is that this agreement with Narozhny and Fedotov exists only when we reach a number of exchanged photons at which the sum-rule \cite{sumrule} is fulfilled. This means that as long as the sum-rule is not achieved at a certain number of exchanged photons, which varies according to the intensities used, the effect of the laser field on the pion lifetime will exist and will take its place. As theoretical physicists, we confirm here that the basis on which this research was built and on which we relied to demonstrate the results obtained is purely theoretical, through which we tried to study theoretically the pion decay and study the behavior of several quantities in the presence of an electromagnetic field, believing that these results may pave the way and offer wider scope for possible experiments in the future. Note that, in this work, natural units $c=\hbar=1$ and the space-time metric $g=diag(1,-1,-1,-1)$ are employed throughout. In many equations of this paper, the Feynman slash notation is used. For any 4-vector, $A$, $\slashed{A}=A^{\mu}\gamma_{\mu}$ where the matrices $\gamma$ are the well-known Dirac matrices. 
This paper is structured as follows. A detailed note on the theoretical model is presented in section II. The results are discussed in section III. Finally, the conclusion of this study is given in section IV.
\section{Outline of the theory}
We consider the decay of a charged pion into two leptons,
\begin{equation}\label{process}
 \pi^{-}(p_{1})\longrightarrow l^{-}(p_{2})+\bar{\nu}_{l}(k'), ~~~~~~(l=e,\mu)
\end{equation}
where $l$ is an electron or a muon and the arguments are our labels for the associated momenta. We assume that this decay occurs in the presence of a circularly polarized monochromatic laser field, which is described by the following classical four-potential:
\begin{align}\label{potential}
A^{\mu}(\phi)=a^{\mu}_{1}\cos(\phi)+a^{\mu}_{2}\sin(\phi),~~~\phi=(k.x),
\end{align}
where $k=(\omega,\textbf{k})$ is the wave 4-vector $(k^{2}=0)$, $\phi$ is the phase of the laser field and $\omega$ its frequency. The polarization 4-vectors $a^{\mu}_{1}$ and $a^{\mu}_{2}$ are equal in magnitude and orthogonal:
\begin{equation}
\begin{split}
a^{\mu}_{1}&=|\text{\textbf{a}}|(0,1,0,0),\\
a^{\mu}_{2}&=|\text{\textbf{a}}|(0,0,1,0),
\end{split}
\end{equation}
which implies $(a_{1}.a_{2})=0$ and $a_{1}^{2}=a_{2}^{2}=a^{2}=-|\text{\textbf{a}}|^{2}=-(\mathcal{E}_{0}/\omega)^{2}$ where $\mathcal{E}_{0}$ is the amplitude of the laser's electric field. We shall assume that the Lorentz gauge condition is applied to the four-potential, so that
\begin{equation}
k_{\mu}A^{\mu}=0,
\end{equation}
which implies $(k.a_{1})=(k.a_{2})=0$, meaning that the wave vector $\textbf{k}$ is chosen to be along the $z$-axis. 
The wave function of the relativistic lepton $l^{-}$ with four-momentum $p_{2}$ moving in an electromagnetic field obeys the following Dirac equation \cite{N}:
\begin{equation}\label{Dirac equation}
\big[(p_{2}-e A)^{2}-m_{l}^{2}-\dfrac{i e}{2}  F_{\mu \nu} \sigma^{\mu \nu}\big]\psi_{l}(x)=0,
\end{equation}
where $e=-|e|$ and $m_{l^{-}}$ are, respectively, the electron charge and the rest mass of the lepton $l^{-}$. Here $F_{\mu \nu}=\partial_{\mu} A_{\nu}-\partial_{\nu} A_{\mu}$ is the electromagnetic field tensor and $\sigma^{\mu \nu}=\tfrac{1}{2}[\gamma^{\mu},\gamma^{\nu}]$.
The solution of Eq.~(\ref{Dirac equation}) gives the relativistic Dirac-Volkov functions \cite{O}, which represent the lepton $l^{-}$ in a laser field normalized to the volume $V$:
\begin{align}\label{muon}
\begin{split}
\psi_{l}(x)=\bigg[1+\dfrac{e\slashed{k}\slashed{A}}{2(k.p_{2})}\bigg]\dfrac{u(p_{2},s_{2})}{\sqrt{2Q_{2}V}}
\times e^{iS(q_{2},x)}
\end{split},
\end{align} 
with 
\begin{equation}
S(q_{2},x)=-q_{2}.x-\dfrac{e(a_{1}.p_{2})}{k.p_{2}}\sin(\phi)+\dfrac{e(a_{2}.p_{2})}{k.p_{2}}\cos(\phi).
\end{equation}
$u(p_{2},s_{2})$ represents the bispinor for the free lepton $l^{-}$ with momentum $p_{2}$ and spin $s_{2}$ satisfying 
\begin{equation}
\sum_{s_{2}}u(p_{2},s_{2})\overline{u}(p_{2},s_{2})=\slashed{p}_{2}+m_{l}.
\end{equation}
The 4-vector $q_{2}=(Q_{2},\textbf{q}_{2})$ is the effective momentum  that the lepton $l^{-}$ acquires in the presence of a
classical monochromatic electromagnetic field
\begin{equation}
q_{2}=p_{2}-\dfrac{e^{2}a^{2}}{2(k.p_{2})}k.
\end{equation}
The square of this four-momentum is given by
\begin{align}
q_{2}^{2}=m_{l}^{2}-e^{2}a^{2}=m_{l*}^{2},
\end{align}
where $m_{l}=0.511~\text{MeV}$ for the electron and $m_{l}=105.6~\text{MeV}$ for the muon.
The quantity $m_{l*}$ plays the role of an effective mass of the lepton $l^{-}$ inside the electromagnetic field.
For the laser-dressed charged pion (spinless particle), its wave function will obey the Klein-Gordon equation for bosons with spin zero, which is in fact the second-order equation (\ref{Dirac equation}) without the term $\tfrac{-i e}{2}F_{\mu \nu}\sigma^{\mu \nu}$. Therefore, the corresponding Volkov solutions read \cite{B}
\begin{align}\label{pion}
\begin{split}
\psi_{\pi^{-}}(x)=\dfrac{1}{\sqrt{2Q_{1}V} }
\times e^{iS(q_{1},x)}
\end{split},
\end{align} 
with
\begin{equation}
S(q_{1},x)=-q_{1}.x-\dfrac{e(a_{1}.p_{1})}{k.p_{1}}\sin(\phi)+\dfrac{e(a_{2}.p_{1})}{k.p_{1}}\cos(\phi).
\end{equation}
The dressed four-momentum $q_{1}=(Q_{1},\textbf{q}_{1})$ and the effective mass $m_{\pi^{-}*}$ of the charged pion are, respectively, such that
\begin{equation}
q_{1}=p_{1}-\dfrac{e^{2}a^{2}}{2(k.p_{1})}k,~~~m_{\pi^{-}*}^{2}=m_{\pi^{-}}^{2}-e^{2}a^{2},
\end{equation}
where $m_{\pi^{-}}=139.57~\text{MeV}$ is the rest mass of the charged pion.
The outgoing antineutrino $\bar{\nu}_{l}$ is treated as massless particle with four-momentum $k'$ and spin $t'$. According to the Feynman rules, it is represented by an incoming wave function with negative four-momentum as follows \cite{J}:
\begin{align}
\psi_{\bar{\nu}_{l}}(x)=\dfrac{v(k',t')}{\sqrt{2E_{2}V}  }e^{ik'.x},
\end{align}
where $E_{2}=k'^{0}$ and $v(k',t')$ is the Dirac spinor satisfying the following formula,
\begin{equation}
\sum_{t'}v(k',t')\overline{v}(k',t')=\slashed{k}'.
\end{equation}
The decay process of the pion in the field of a circularly polarized electromagnetic plane wave is a weak interaction process; it can be described by the lowest Feynman diagrams. Therefore,  in the first Born approximation, the S-matrix   element for the laser-assisted   $\pi^{-}$ decay can be written as \cite{J}
\begin{align}\label{smatrix}
S_{fi}(\pi^{-}\rightarrow l^{-}\bar{\nu}_{l})=\dfrac{-iG}{\sqrt{2}}\int d^{4}xJ^{(\pi)\dagger}_{\mu}(x)J^{\mu}_{(l^{-})}(x).
\end{align}
Here $G=(1.166~37\pm0.000~02)\times10^{-11}~\text{MeV}^{-2}$ is the Fermi coupling constant, $J^{\mu}_{(l^{-})}(x)$ and $J^{(\pi)}_{\mu}(x)$ are, respectively, the leptonic and hadronic currents in the laser field, which can be expressed by
\begin{align}\label{currlepton}
J^{\mu}_{(l^{-})}(x)=\bar{\psi}_{l}(x,t)\gamma^{\mu}(1-\gamma_{5})\psi_{\bar{\nu}_{l}}(x,t),
\end{align}
and 
\begin{align}\label{currpion}
J^{(\pi)}_{\mu}=i\sqrt{2}f_{\pi}p_{1\mu}\dfrac{1}{\sqrt{2Q_{1}V} }
\times e^{-iS(q_{1},x)},
\end{align}
where $f_{\pi}=90.8~\text{MeV}$ is called commonly the pion decay constant \cite{J}. We note that the sign of the argument of the exponential function in (\ref{currpion}) is chosen in such a way that the product of all plane waves in the S-matrix element yields the conservation of four-momentum. This corresponds to assigning the character of an antiparticle to the negative pion, while the positive pion has the character of a particle \cite{J}.
Inserting Eqs.~(\ref{currlepton}) and (\ref{currpion}) and the wave functions into Eq.~(\ref{smatrix}) and after some manipulations, we find the S-matrix to be written as
\begin{align}\label{smatrix1}
\begin{split}
S_{fi}&=\dfrac{-Gf_{\pi}}{2\sqrt{2Q_{1}Q_{2}E_{2}V^{3}}}\int d^{4}xp_{1\mu}\bar{u}(p_{2},s_{2})\big[1+ C(p_{2})\slashed{a}_{1}\slashed{k}\cos(\phi)+C(p_{2})\slashed{a}_{2}\slashed{k}\sin(\phi)\big]\gamma^{\mu}(1-\gamma_{5})v(k',t')\\
& ~~~~~~~~~~~~~~~~~~~~~~~~~~~~~~~~~~~~~~~ \times e^{ik'.x}~ e^{i(S(q_{1},x)-S(q_{2},x))},
\end{split}
\end{align}
where $C(p_{2})=e/(2 (k.p_{2}))$. Now, we transform $e^{i(S(q_{1},x)-S(q_{2},x))}$ by introducing
\begin{align}\label{argument}
z=\sqrt{\alpha_{1}^{2}+\alpha_{2}^{2}}\quad\text{with}\quad\alpha_{1}=e\bigg(\dfrac{a_{1}.p_{1}}{k.p_{1}}-\dfrac{a_{1}.p_{2}}{k.p_{2}}\bigg)\,;\,\alpha_{2}=e\bigg(\dfrac{a_{2}.p_{1}}{k.p{1}}-\dfrac{a_{2}.p_{2}}{k.p_{2}}\bigg),
\end{align}
we get
\begin{align}
S(q_{1},x)-S(q_{2},x)=(q_{2}-q_{1})x-z\sin(\phi-\phi_{0}),
\end{align}
with $\phi_{0}=\atan(\alpha_{2}/\alpha_{1})$. Therefore, the S-matrix element becomes
\begin{align}\label{smatrix2}
\begin{split}
S_{fi}&=\dfrac{-Gf_{\pi}}{2\sqrt{2Q_{1}Q_{2}E_{2}V^{3}}}\int d^{4}xp_{1\mu}\bar{u}(p_{2},s_{2})\big[1+ C(p_{2})\slashed{a}_{1}\slashed{k}\cos(\phi)+C(p_{2})\slashed{a}_{2}\slashed{k}\sin(\phi)\big]\gamma^{\mu}(1-\gamma_{5})v(k',t')\\
& ~~~~~~~~~~~~~~~~~~~~~~~~~~~~~~~~~~~~~~~ \times e^{i(k'+q_{2}-q_{1}).x}~ e^{-iz\sin(\phi-\phi_{0})}.
\end{split}
\end{align}
The three  different  quantities  in Eq.~(\ref{smatrix2})  can be transformed by the well-known identities involving ordinary Bessel functions  $J_{s}(z)$:
\begin{align}\label{transformation}
\begin{bmatrix}
1\\
\cos(\phi)\\
\sin(\phi)
\end{bmatrix}\times e^{-iz\sin(\phi-\phi_{0})}=\sum_{s=-\infty}^{+\infty}\begin{bmatrix}
B_{s}(z)\\
B_{1s}(z)\\
B_{2s}(z)
\end{bmatrix}e^{-is\phi},
\end{align}
where
\begin{align}
\begin{split}
\begin{bmatrix}
B_{s}(z)\\
B_{1s}(z)\\
B_{2s}(z) \end{bmatrix}=\begin{bmatrix}J_{s}(z)e^{is\phi_{0}}\\
\big(J_{s+1}(z)e^{i(s+1)\phi_{0}}+J_{s-1}(z)e^{i(s-1)\phi_{0}}\big)/2\\
\big(J_{s+1}(z)e^{i(s+1)\phi_{0}}-J_{s-1}(z)e^{i(s-1)\phi_{0}}\big)/2i
 \end{bmatrix},
\end{split}
\end{align}
where $z$ is the argument of the Bessel functions defined in Eq.~(\ref{argument}) and $s$  is the number of exchanged photons. Using these transformations in Eq.~(\ref{smatrix2}) and  integrating over $d^{4}x$, the matrix element $S_{fi}$ becomes
\begin{align}
S_{fi}=\dfrac{-Gf_{\pi}}{2\sqrt{2Q_{1}Q_{2}E_{2}V^{3}}}\sum_{s=-\infty}^{\infty}M^{s}_{fi}(2\pi)^{4}\delta^{4}(k'+q_{2}-q_{1}-sk),
\end{align}
where the quantity $M^{s}_{fi}$ is defined by
\begin{align}
M^{s}_{fi}=\bar{u}(p_{2},s_{2})\Gamma^{s}v(k',t'),
\end{align}
where
\begin{align}
\Gamma^{s}=\big[B_{s}(z)+ C(p_{2}) \slashed{a}_{1}\slashed{k} B_{1s}(z)+C(p_{2}) \slashed{a}_{2}\slashed{k} B_{2s}(z)\big]\slashed{p}_{1}(1-\gamma_{5}).
\end{align}
To evaluate the pion lifetime, we first evaluate the decay rate of the pion per particle and per time into the final states, which are obtained by multiplying the squared S-matrix element by the density of final states, summing over spins of leptons and antineutrinos and finally dividing by the time $\text{T}$. We obtain for the decay rate of the pion:
\begin{align}\label{summed}
W(\pi^{-}\rightarrow l^{-}\bar{\nu}_{l})=\sum_{s=-\infty}^{+\infty}W_{s}(\pi^{-}\rightarrow l^{-}\bar{\nu}_{l}),
\end{align}
where the photon-number-resolved decay rate $W_{s}$ is defined by
 \begin{align}
W_{s}(\pi^{-}\longrightarrow l^{-}+\bar{\nu}_{l})=\dfrac{G^{2}f_{\pi}^{2}}{8Q_{1}}\int\dfrac{d^{3}q_{2}}{(2\pi)^{3}Q_{2}}\int\dfrac{d^{3}k'}{(2\pi)^{3}E_{2}}(2\pi)^{4}\delta^{4}(k'+q_{2}-q_{1}-sk)|\overline{M^{s}_{fi}}|^{2},
\end{align}
where
\begin{align}
|\overline{M^{s}_{fi}}|^{2}=\sum_{s_{2},t'}|M^{s}_{fi}|^{2}=\sum_{s_{2},t'}|\bar{u}(p_{2},s_{2})\Gamma^{s}v(k',t')|^{2}.
\end{align}
Let us recall that the measured quantity here is the decay rate $W(\pi^{-}\rightarrow l^{-}\bar{\nu}_{l})$ obtained from the so-called Breit-Wigner distribution, which represents the measurement of the invariant mass of unstable particle.\\
Performing the integration over $d^{3}k'$, the photon-number-resolved decay rate $W_{s}$ becomes
\begin{align}\label{28}
W_{s}=\dfrac{G^{2}f_{\pi}^{2}}{(2\pi)^{2}8Q_{1}}\int\dfrac{d^{3}q_{2}}{E_{2}Q_{2}}\delta(E_{2}+Q_{2}-Q_{1}-s\omega)|\overline{M^{s}_{fi}}|^{2},
\end{align}
with $\textbf{k}'+\textbf{q}_{2}-\textbf{q}_{1}-s\textbf{k}=0$. In the pion rest frame, we furthermore have $Q_{1}=m_{\pi^{-}*}$ and $\textbf{q}_{1}=0$, then $\textbf{k}'=s\textbf{k}-\textbf{q}_{2}$. Hence, using $ d^{3}q_{2}=|\textbf{q}_{2}|^{2}d|\textbf{q}_{2}|d\Omega_{l}$, we obtain
\begin{align}\label{29}
W_{s}=\dfrac{G^{2}f_{\pi}^{2}}{(2\pi)^{2}8Q_{1}}\int \dfrac{|\textbf{q}_{2}|^{2}d|\textbf{q}_{2}|d\Omega_{l}}{E_{2}Q_{2}}~\delta\Big(\sqrt{(s\omega)^{2}+|\textbf{q}_{2}|^{2}-2s\omega|\textbf{q}_{2}|\cos(\theta)}+\sqrt{|\textbf{q}_{2}|^{2}+m_{l*}^{2}}-Q_{1}-s\omega\Big)|\overline{M^{s}_{fi}}|^{2}.
\end{align}
The remaining integral over $d|\textbf{q}_{2}|$ can be solved by using the familiar formula:
\begin{align}\label{familiarformula}
\int dxf(x)\delta(g(x))=\dfrac{f(x)}{|g'(x)|}\bigg|_{g(x)=0}.
\end{align}
Thus we get
\begin{align}\label{ws}
W_{s}=\dfrac{G^{2}f_{\pi}^{2}}{(2\pi)^{2}8Q_{1}}\int \dfrac{|\textbf{q}_{2}|^{2}d\Omega_{l}}{E_{2}Q_{2}g'(|\textbf{q}_{2}|)}|\overline{M^{s}_{fi}}|^{2},
\end{align}
where
\begin{align}
g'(|\textbf{q}_{2}|)=\dfrac{|\textbf{q}_{2}|-s\omega\cos(\theta)}{\sqrt{(s\omega)^{2}+|\textbf{q}_{2}|^{2}-2s\omega|\textbf{q}_{2}|\cos(\theta)}}+\dfrac{|\textbf{q}_{2}|}{\sqrt{|\textbf{q}_{2}|^{2}+m_{l*}^{2}}}.
\end{align}
The term  $|\overline{M^{s}_{fi}}|^{2}$ can be calculated as  follows:
\begin{align}\label{trace}
|\overline{M^{s}_{fi}}|^{2}=\text{Tr}\big[(\slashed{p}_{2}+m_{l})\Gamma^{s}\slashed{k}'\bar{\Gamma}^{s}\big],
\end{align}
where
\begin{align}
\begin{split}
\bar{\Gamma}^{s}&=\gamma^{0}\Gamma^{s\dagger}\gamma^{0},\\
&=\slashed{p}_{1}\big(1-\gamma_{5}\big)\big[B^{*}_{s}(z)+ C(p_{2})\slashed{k}\slashed{a}_{1}B^{*}_{1s}(z)+C(p_{2}) \slashed{k}\slashed{a}_{2}B^{*}_{2s}(z)\big].
\end{split}
\end{align}
The trace calculation can be performed with the help of FEYNCALC \cite{P}. The result of the trace (\ref{trace}) is given by
\begin{align}\label{result}
|\overline{M^{s}_{fi}}|^{2}=\dfrac{2e^{-i(s+2)\phi_{0}}}{(k.p_{2})}\bigg[AJ_{s-1}^{2}(z)+BJ_{s+1}^{2}(z)+CJ_{s-1}(z)J_{s}(z)+DJ_{s+1}(z)J_{s}(z)+EJ_{s}^{2}(z)\bigg],
\end{align}
where the five coefficients $A$, $B$, $C$, $D$ and $E$ are explicitly given by
\begin{align}
A=-e^{2}e^{i(s+2)\phi_{0}}\big[m_{\pi}^{2}\epsilon(a_{1},a_{2},k ,k')-2 ~(k'.p_{1})\epsilon(a_{1},a_{2},k,p_{1})+a^{2}\big(2(k'.p_{1})( k.p_{1})-(k.k')m_{\pi}^{2}\big)\big],
\end{align}
\begin{align}
B=-e^{2}e^{i(s+2)\phi_{0}}\big[-m_{\pi}^{2} \epsilon(a_{1},a_{2},k ,k')+2(k'.p_{1})\epsilon(a_{1},a_{2},k,p_{1})+a^{2}\big(2(k'.p_{1})(k.p_{1})-(k.k')m_{\pi}^{2}\big)\big],
\end{align}
\begin{align}
\begin{split}
C&=e\big\lbrace im_{\pi}^{2}\big(e^{i(s+1)\phi_{0}}-e^{i(s+3)\phi_{0}}\big)\epsilon(a_{1},k,k',p_{2})-2i(k'.p_{1})\big(e^{i(s+1)\phi_{0}}-e^{i(s+3)\phi_{0}}\big)\\
&\times \epsilon(a_{1},k,p_{1} ,p_{2})+e^{i(s+1)\phi_{0}}\big[-m_{\pi}^{2}\big(1+e^{2i\phi_{0}}\big)\epsilon(a_{2},k,k',p_{2})+2(k'.p_{1})\big(1+e^{2i\phi_{0}}\big)\\
&\times \epsilon(a_{2},k,p_{1} ,p_{2})+(a_{1}.k')( k.p_{2})m_{\pi}^{2}e^{2i\phi_{0}}+(a_{1}.k')( k.p_{2})m_{\pi}^{2}-(a_{1}.p_{2})(k.k')m_{\pi}^{2} e^{2i\phi_{0}}\\
&-(a_{1}.p_{2})(k.k')m_{\pi}^{2}+2(a_{1}.p_{2})(k'.p_{1})(k.p_{1})e^{2i\phi_{0}}+2(a_{1}.p_{2})(k'.p_{1})(k.p_{1})-i(a_{2}.k')(k.p_{2})\\ 
&\times m_{\pi}^{2}e^{2i\phi_{0}}+i(a_{2}.k')(k.p_{2})m_{\pi}^{2}+i(a_{2}.p_{2})( k.k')m_{\pi}^{2}e^{2i\phi_{0}}-i (a_{2}.p_{2})(k.k')m_{\pi}^{2}\\
&-2i(a_{2}.p_{2})(k'.p_{1})(k.p_{1})e^{2i\phi_{0}}+2i(a_{2}.p_{2})( k'.p_{1})( k.p_{1})\big]\big\rbrace,\\
\end{split}
\end{align}
\begin{align}
\begin{split}
D&=e\big\lbrace -im_{\pi}^{2}\big(e^{i(s+1)\phi_{0}}-e^{i(s+3)\phi_{0}}\big)\epsilon(a_{1},k,k',p_{2})+2i(k'.p_{1})\big(e^{i(s+1)\phi_{0}}-e^{i(s+3)\phi_{0}}\big)\\
&\times \epsilon(a_{1},k,p_{1},p_{2})+e^{i(s+1)\phi_{0}}\big[m_{\pi}^{2}\big(1+e^{2i\phi_{0}}\big)\epsilon(a_{2},k,k',p_{2})-2(k'.p_{1})\big(1+e^{2i\phi_{0}}\big)\\
& \times \epsilon(a_{2},k,p_{1},p_{2})+(a_{1}.k')( k.p_{2})m_{\pi}^{2}e^{2i\phi_{0}}+(a_{1}.k')( k.p_{2})~m_{\pi}^{2}-(a_{1}.p_{2})(k.k')m_{\pi}^{2}e^{2i\phi_{0}}\\
&-(a_{1}.p_{2})(k.k')m_{\pi}^{2}+2(a_{1}.p_{2})(k'.p_{1})(k.p_{1})e^{2i\phi_{0}}+2(a_{1}.p_{2})(k'.p_{1})(k.p_{1})-i(a_{2}.k')(k.p_{2})\\
&\times m_{\pi}^{2}e^{2i\phi_{0}}+i(a_{2}.k')(k.p_{2})m_{\pi}^{2}+i(a_{2}.p_{2})(k.k')m_{\pi}^{2}e^{2i\phi_{0}}-i(a_{2}.p_{2})(k.k')m_{\pi}^{2}\\
&-2i(a_{2}.p_{2})(k'.p_{1})(k.p_{1})e^{2i\phi_{0}}+2i(a_{2}.p_{2})(k'p_{1})(k.p_{1})\big]\big\rbrace,
\end{split}
\end{align}
\begin{align}
E=-4(k.p_{2})e^{i(s+2)\phi_{0}}\big[(k'.p_{2})m_{\pi}^{2}-2(k'.p_{1})(p_{1}.p_{2})\big],
\end{align}
where, for all 4-vectors $a, b, c$ and $d$, we have
\begin{equation}
\epsilon(a,b,c,d)=\epsilon_{\mu\nu\rho\sigma}a^{\mu}b^{\nu}c^{\rho}d^{\sigma}.
\end{equation}
We notice that in the coefficient $E$, there is no occurrence of the antisymmetric tensors $\epsilon_{\mu\nu\rho\sigma}$. This clearly means that they were totally contracted. The other coefficients $A$, $B$, $C$ and $D$ contained various non contracted tensors. For example in $A$ and $B$, there are two non contracted tensors involving $\epsilon_{\mu\nu\rho\sigma}$ whereas in $C$ and $D$, there are four. Particle physicists are very often dealing with this. Let us remind that to evaluate these tensors, we use the Grozin convention
\begin{equation}
\epsilon_{0123}=1
\end{equation}
meaning that $\epsilon_{\mu\nu\rho\sigma}=1$ for an even permutation of the Lorentz indices whereas $\epsilon_{\mu\nu\rho\sigma}=-1$ for an odd permutation of the Lorentz indices and finally $\epsilon_{\mu\nu\rho\sigma}=0$ otherwise. Using Einstein's summation convention, the non contracted tensors in $A$ and $B$ reduce to
\begin{equation}
\begin{split}
\epsilon(a_{1},a_{2},k,k')&=\epsilon_{\mu\nu\rho\sigma}a_{1}^{\mu}a_{2}^{\nu}k^{\rho}k'^{\sigma},\\
&=|\text{\textbf{a}}|^{2}\big[\epsilon_{1203}k^{0}k'^{3}+\epsilon_{1230}k^{3}k'^{0}\big],\\
&=|\text{\textbf{a}}|^{2}\omega\big[s\omega-|\textbf{q}_{2}|\cos(\theta)-E_{2}\big],
\end{split}
\end{equation}
and
\begin{equation}
\begin{split}
\epsilon(a_{1},a_{2},k,p_{1})&=\epsilon_{\mu\nu\rho\sigma}a_{1}^{\mu}a_{2}^{\nu}k^{\rho}p_{1}^{\sigma},\\
&=|\text{\textbf{a}}|^{2}\big[\epsilon_{1203}k^{0}p_{1}^{3}+\epsilon_{1230}k^{3}p_{1}^{0}\big],\\
&=|\text{\textbf{a}}|^{2}\omega\bigg[\frac{e^{2}a^{2}\omega}{2(k.p_{1})}-p_{1}^{0}\bigg].
\end{split}
\end{equation}
With the same approach, the four tensors appearing in $C$ and $D$ can be expressed as follows:
\begin{equation}
\begin{split}
\epsilon(a_{1},k,k',p_{2})&=|\text{\textbf{a}}||\textbf{q}_{2}|\omega\sin(\theta)\sin(\varphi)\bigg[s\omega+\frac{e^{2}a^{2}\omega}{2(k.p_{2})}-E_{2}-p_{2}^{0}\bigg],\\
\epsilon(a_{1},k,p_{1},p_{2})&=|\text{\textbf{a}}||\textbf{q}_{2}|\omega \sin(\theta)\sin(\varphi)\bigg[\frac{e^{2}a^{2}\omega}{2(k.p_{1})}-p_{1}^{0}\bigg],\\
\epsilon(a_{2},k,k',p_{2})&=|\text{\textbf{a}}||\textbf{q}_{2}|\omega \sin(\theta)\cos(\varphi)\bigg[E_{2}+p_{2}^{0}-s\omega-\frac{e^{2}a^{2}\omega}{2(k.p_{2})}\bigg],\\
\epsilon(a_{2},k,p_{1},p_{2})&=|\text{\textbf{a}}||\textbf{q}_{2}|\omega \sin(\theta)\cos(\varphi)\bigg[p_{1}^{0}-\frac{e^{2}a^{2}\omega}{2(k.p_{1})}\bigg],
\end{split}
\end{equation}
where $\theta$ and $\varphi$ are the spherical coordinates of $\textbf{q}_{2}$. $p_{1}^{0}$ and $p_{2}^{0}$ are, respectively, the temporal components of $p_{1}$ and $p_{2}$ given by
\begin{equation}
\begin{split}
p_{1}^{0}=Q_{1}+\frac{e^{2}a^{2}\omega}{2(k.p_{1})}\quad;\quad
p_{2}^{0}=Q_{2}+\frac{e^{2}a^{2}\omega}{2(k.p_{2})}.
\end{split}
\end{equation}
After giving some important details about the trace calculation, we return to the lifetime of the charged pion, which is defined by
\begin{align}
\tau_{\pi^{-}}=\dfrac{1}{W_{total}},
\end{align}
where $W_{total}=W(\pi^{-}\rightarrow \mu^{-}\bar{\nu}_{\mu})+W(\pi^{-}\rightarrow e^{-}\bar{\nu}_{e})$ is the total decay rate of the charged pion $\pi^{-}$ in the laser field. Now, we introduce a very interesting quantity measured experimentally. This is the branching ratio (Br) of a decay mode. In particle physics, the branching ratio refers to the probability that a particle will follow a given decay mode out of all possible decay modes. The sum of the branching ratios of all decay modes of a particle is therefore by definition equal to $1$ (or $100\%$). 
In our case, we define the branching ratios (Brs) of the muonic and electronic decay channels as follows:
\begin{eqnarray}
\text{Br}(\pi^{-}\rightarrow \mu^{-}\bar{\nu}_{\mu})&=&\frac{W(\pi^{-}\rightarrow \mu^{-}\bar{\nu}_{\mu})}{W_{total}},\label{brmuon}\\
\text{Br}(\pi^{-}\rightarrow e^{-}\bar{\nu}_{e})&=&\frac{W(\pi^{-}\rightarrow e^{-}\bar{\nu}_{e})}{W_{total}}\label{brelec}.
\end{eqnarray}
The ratio between the two decay channels, also called the branching ratio for the decay modes, is given by
\begin{equation}\label{ratioexp}
R_{e/ \mu}=\frac{W(\pi^{-}\rightarrow e^{-}\bar{\nu}_{e})}{W(\pi^{-}\rightarrow \mu^{-}\bar{\nu}_{\mu})}.
\end{equation}
Its experimental value, in the absence of the laser field, is $R^{exp}_{e/ \mu}=(1.218\pm0.014)\times10^{-4}$ \cite{Brexp}.
\section{Results and Discussion}
In this section, we present and analyze the numerical results for the pion decay  in the presence of a circularly polarized laser field. It is important here to warn that all obtained numerical results are only a means of depicting the theoretical quantities that were calculated in the previous section. The influence of the laser parameters (intensity and frequency) on the decay rate, lifetime, and branching ratio is discussed. The origin of the coordinate system is chosen to be on the pion which is at rest before decay. The direction of the field wave vector $\textbf{k}$ is along the $z$-axis. The integral over $d\Omega_{l}$ ($d\Omega_{l}=\sin(\theta)d\theta d\varphi$) involved in the evaluation of $W_{s}$ (\ref{ws}) should be performed using the numerical integration. The spherical coordinate $\varphi$ is chosen to be $\varphi=0^{\circ}$ throughout this section. The expression of $|\textbf{q}_{2}|$ can be found by solving the equation $g(|\textbf{q}_{2}|)=0$, which is a condition to apply the familiar formula (\ref{familiarformula}). The relation between the lifetime $\tau$ expressed in seconds [$\text{s}$] and the decay rate $W$ expressed in [$\text{eV}$] can be obtained from
\begin{equation}
\tau~[\text{s}]=\frac{6.58212\times 10^{-16}~[\text{eV.s}]}{W~[\text{eV}]}.
\end{equation} 
Bearing in mind that we have used the following unit conversion: in System International (SI) units, the electric field strength $\mathcal{E}_{0}[SI]=4329.0844~[\text{V/cm}]$ corresponds, in natural units, to $\mathcal{E}_{0}[NU]=1~[\text{eV}^{2}]$.
\begin{figure}[hptb]
\subfloat[]{\label{a}\includegraphics[height=5cm,width=.45\linewidth]{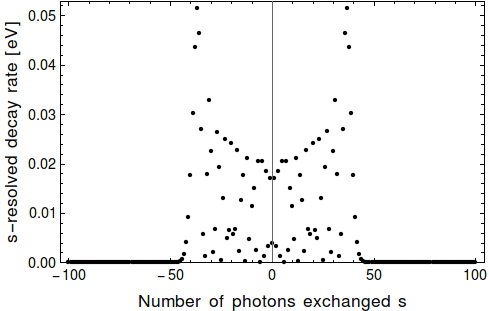}}\hspace*{0.65cm}
\subfloat[]{\label{b}\includegraphics[height=5cm,width=.45\linewidth]{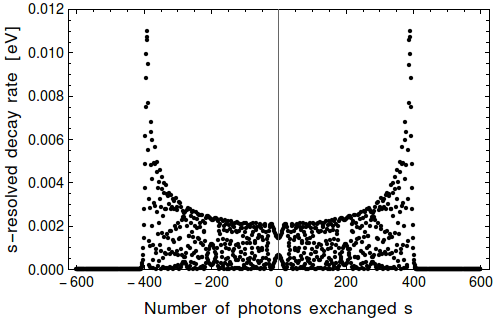}}\par 
\subfloat[]{\label{c}\includegraphics[height=5cm,width=.45\linewidth]{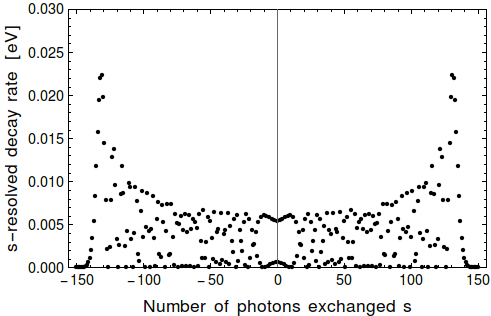}}
\caption{The multi-photon decay rate $W_{s}(\pi^{-}\rightarrow \mu^{-}\bar{\nu}_{\mu})$ (\ref{ws}) (in units of $10^{-8}$) as a function of the number of photons exchanged $s$ in the rest frame of the pion, with the spherical coordinates $\theta=90^{\circ}$ and $\varphi=0^{\circ}$. The laser field amplitude and frequency are (a) $\mathcal{E}_{0}=10^{7}~\text{V/cm}$ and $\hbar\omega=1.17~\text{eV}$, (b) $\mathcal{E}_{0}=10^{6}~\text{V/cm}$ and $\hbar\omega=0.117~\text{eV}$ or $\mathcal{E}_{0}=10^{8}~\text{V/cm}$ and $\hbar\omega=1.17~\text{eV}$, and (c) $\mathcal{E}_{0}=10^{8}~\text{V/cm}$ and $\hbar\omega=2~\text{eV}$.}
\label{fig1}
\end{figure}
In the following and throughout this section, we will adopt the same arrangement that was followed to construct the theory in the previous section. We will start by showing the results of the data obtained for the decay rate, then for the lifetime and finally for the branching ratio. We will study the effect of the laser field on each of these quantities. For the results related to the decay rate, we remind here that only the muonic channel $(\pi^{-}\rightarrow \mu^{-}\bar{\nu}_{\mu})$ has been considered, since it is more favored than the electronic channel for considerations and reasons that will be mentioned at the end of this section.
Figure \ref{fig1} shows the phenomenon of multi-photon energy transfer for different field strengths and frequencies. In Fig.~\ref{a}, we display the photon-number-resolved decay rate $W_{s}(\pi^{-}\rightarrow \mu^{-}\bar{\nu}_{\mu})$ (\ref{ws}) versus the net photon number $s$ transferred between the decaying system and the laser field. We have chosen the spherical coordinates $\theta=90^{\circ}$ and $\varphi=0^{\circ}$. The strength and frequency of the laser field are $\mathcal{E}_{0}=10^{7}~\text{V/cm}$ and $\hbar\omega=1.17~\text{eV}$. A few number of photons are exchanged between the laser field and the decaying system. The cutoffs are $s\simeq -50$ photons for the negative part (absorption) of the envelope and $s\simeq +50$ photons for the positive part (emission). Figure \ref{b} shows the situation when only the field strength increases to $10^{8}~\text{V/cm}$ and also when both the field strength and frequency decrease, respectively, to $10^{6}~\text{V/cm}$ and $\hbar\omega=0.117~\text{eV}$. The other parameters are the same as in Fig.~\ref{a}. The process involving large numbers of photons exchanged has significant contribution and the cutoff numbers now are about $s=-450$ and $s=+450$. Comparisons between Figs.~\ref{a} and \ref{b} show that the transfer of photons is enhanced when only the strength of the laser field is increased. Figure \ref{c} shows the behavior of $W_{s}(\pi^{-}\rightarrow \mu^{-}\bar{\nu}_{\mu})$ when $\mathcal{E}_{0}=10^{8}~\text{V/cm}$ and $\hbar\omega=2~\text{eV}$ and the cutoff numbers, in this case, are $s=-150$ and $s=+150$. In Fig.~\ref{fig1}, the contributions of various $s$-photon processes are cut off at two edges which are symmetric with respect to $s=0$. The spectrum exhibits also an overall symmetric envelope for peaks of negative and  positive energy transfer. The cutoff number can be explained by the properties of the Bessel function, which decreases exponentially after its order reaches its argument. The heights of the different photon-energy-transfer peaks depend crucially on the values of the ordinary Bessel functions. 
\begin{figure}[hptb]
\subfloat[]{\label{a2}\includegraphics[height=5cm,width=.45\linewidth]{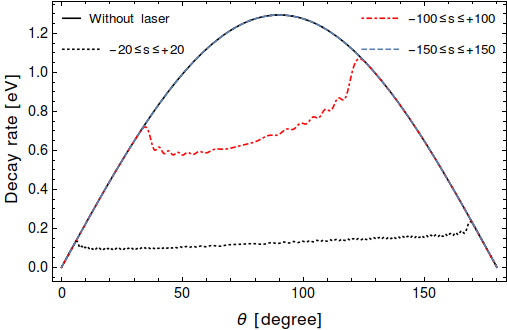}}\hspace*{.5cm}
\subfloat[]{\label{b2}\includegraphics[height=5cm,width=.45\linewidth]{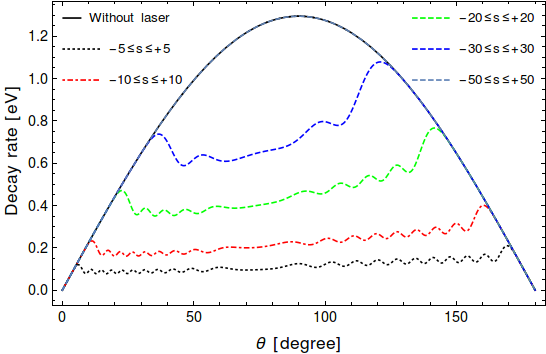}}
\caption{The variation of the summed decay rates $W(\pi^{-}\rightarrow \mu^{-}\bar{\nu}_{\mu})$ (\ref{summed}) (in units of $10^{-8}$) with and without a laser as a function of the angle $\theta$ for various numbers of photons exchanged. The spherical coordinate $\varphi=0^{\circ}$. The laser field amplitude and frequency are (a) $\mathcal{E}_{0}=10^{8}~\text{V/cm}$ and $\hbar\omega=2~\text{eV}$, (b) $\mathcal{E}_{0}=10^{7}~\text{V/cm}$ and $\hbar\omega=1.17~\text{eV}$.}
\label{fig2}
\end{figure}\\
In Fig.~\ref{fig2}, we have made simulations concerning the laser-assisted decay rate for a set of net number of photons exchanged. In Figs.~\ref{a2} and \ref{b2}, we have summed, respectively, over these sets ($-N\leq s\leq +N$ with $N=20,~100,~150$) and ($-N\leq s\leq +N$ with $N=5,~10,~20,~30,~50$) for different field strengths and frequencies. We see that, in Fig.~\ref{a2} (Fig.~\ref{b2}), at $-150\leq s\leq +150$ ($-50\leq s\leq +50$) the two decay rates with and without laser field give two indistinguishable curves. Otherwise, the laser field gives rise to significant changes in the decay rate. We return to the case $-150\leq s\leq +150$ in Fig.~\ref{a2}; the coincidence reached here is called the sum-rule that was shown by Bunkin and Fedorov as well as by Kroll and Watson \cite{sumrule}. Let us first recall that the sum-rule is, mathematically, modeled such as
\begin{equation}
\sum_{s=-\infty}^{+\infty} W^{s}(\pi^{-}\rightarrow l^{-}\bar{\nu}_{l})=W^{laser-free}(\pi^{-}\rightarrow l^{-}\bar{\nu}_{l}),
\end{equation}
and is achieved when the laser-assisted decay rate tends to approach
the laser-free results with increasing the number of photons exchanged. Figures \ref{c} and \ref{a2} perfectly establish correlation in the net number of photons exchanged that reaches the well-known sum-rule. As we can see from Fig.~\ref{c}, the decay rate falls off abruptly beyond the interval $[-150,~+150]$ and Fig.~\ref{a2} shows clearly that beyond $-150\leq s\leq +150$, the sum-rule is obviously checked. The same thing can be said in the case of Figs.~\ref{b2} and \ref{a} in which the sum-rule is attained only at $-50\leq s\leq +50$.
After extensive discussion regarding the effect of the laser field on the decay rate, let us move on to discuss the influence of the laser on the pion lifetime as an important point in our research. 
\begin{figure}[hbtp]
 \centering
 \includegraphics[scale=0.55]{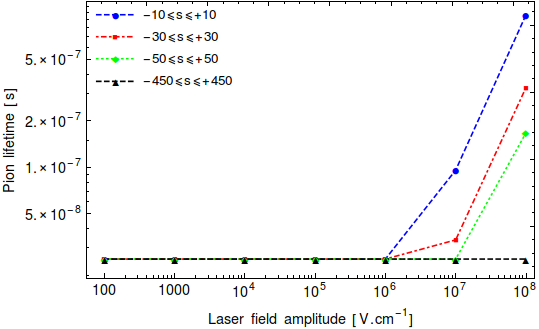}
 \caption{The laser-modified pion lifetime as a
function of the laser field amplitude for various numbers of photons exchanged. The frequency of the laser field is $\hbar\omega=1.17~\text{eV}$.}\label{lifetime1}
\end{figure}\\
Figure \ref{lifetime1} displays the typical behavior of the pion lifetime in the rest frame of the pion for the laser frequency $\hbar\omega=1.17~\text{eV}$ and for different numbers of photons exchanged. As we can see from this figure,  at small intensities $(10^{2}-10^{6}~\text{V/cm})$ all curves, regardless of the number of photons exchanged, are identical and take a fixed value equal to the value of the pion lifetime in the absence of the laser field. Beyond the intensity $10^{6}~\text{V/cm}$, we notice that the lifetime, for each specified number of exchanged photons, expands and lengthens (i.e. it is subject to elongation and extension) and changes nonlinearly with increasing intensity. It seems clear to us that as the number of exchanged photons increases, the effect of the laser on the lifetime diminishes until it becomes zero when we reach $-450\leq s\leq +450$ exchanged photons at which the well-known sum-rule is achieved. 
  \begin{figure}[hbtp]
 \centering
 \includegraphics[scale=0.55]{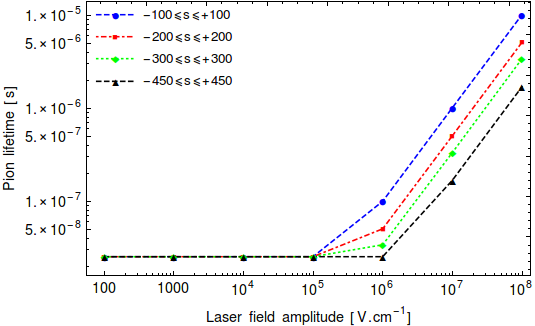}
 \caption{The laser-modified pion lifetime as a
function of the laser field amplitude for various numbers of photons exchanged. The frequency of the laser field is $\hbar\omega=0.117~\text{eV}$.}\label{lifetime2}
\end{figure}\\
Figure \ref{lifetime2} represents the same thing as shown in Fig.~\ref{lifetime1}, but here the laser frequency is equal to $\hbar\omega=0.117~\text{eV}$. We note from this figure that, in the interval of intensities $(10^{2}-10^{5}~\text{V/cm})$, the laser has no effect on the lifetime, regardless of the number of photons exchanged. However, between the intensities $10^{5}$ and $10^{6}~\text{V/cm}$, the effect of the laser on the lifetime appears only at $-100\leq s\leq +100$, $-200\leq s\leq +200$ and $-300\leq s\leq +300$, and the laser field has no effect at all on the case of the number of exchanged photons $-450\leq s\leq +450$, since the latter presents the cutoff number in the case of intensity $10^{6}~\text{V/cm}$ and frequency $\omega=0.117~\text{eV}$ as indicated in Fig.~\ref{b}. As expected, in the interval of intensities $(10^{6}-10^{8}~\text{V/cm})$, the effect of the laser on the lifetime will come back as long as we do not reach a number of exchanged photons at which the sum-rule is achieved. For the intensity $10^{8}~\text{V/cm}$ and the frequency $\omega=0.117~\text{eV}$, the cutoffs are at $s=-50000$ and $s=+50000$. We point out here that due to our limited computational computing, we could not include the result obtained with respect to this large number of exchanged photons. Nevertheless, we are certain that, when we reach $50000$  photons exchanged, the effect of the laser on the lifetime will not remain, and we will obtain a fixed horizontal curve at the intensities between $10^{2}$ and $10^{8}~\text{V/cm}$. For the effect of the laser frequency, and by comparison between Figs. \ref{lifetime1} and \ref{lifetime2}, it becomes clear to us that at higher frequencies, the effect of the laser on the lifetime decreases, a behavior similar to the case of the muon lifetime \cite{D}. Besides all this, we confirm that this observed change in the pion lifetime is not strange or contradictory to physical intuition. Furthermore, a long lifetime can be easily understood as a result of the so-called Zeno effect, which has received great attention from many scientific research in the past and present \cite{Zenoeffect}. In $1977$, Misra and Sudarshan \cite{firstZeno} showed, based on the quantum measurement theory, that frequent observations slow down the decay and may alter the decay rate. They were the first to call the effect by that name. It was predicted that an unstable particle would never even decay when continuously observed. In the paper \cite{Hindered}, it was shown that the lifetime of an unstable system can be extended by watching it closely (e.g., illuminating it with an intense laser field of appropriate frequency). Consequently, we attribute this observed change in lifetime to the quantum Zeno effect resulting from the interaction of the decaying system with the external electromagnetic field which, in our case, plays the role of a measurement device. 
\begin{figure}[hbtp]
\centering
\includegraphics[scale=0.6]{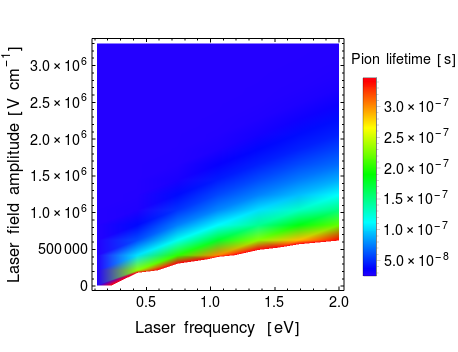}
\caption{The behavior of the pion lifetime as a function of the laser field amplitude $\mathcal{E}_{0}$ and laser frequency $\omega$ for an exchange of $-5\leq s\leq +5$ photons. The laser field amplitude varies from $10^{4}~\text{V/cm}$ to $3.3\times 10^{6}~\text{V/cm}$ whereas the frequency varies from $0.117$ to $2~\text{eV}$.}\label{lifetime3d}
\end{figure}
To show the global behavior of the pion lifetime, we simultaneously vary  the laser intensity and frequency. Figure \ref{lifetime3d} illustrates the density plot for the pion lifetime over the $(\omega,\mathcal{E}_{0})$ plane for $-5\leq s\leq +5$ photons exchanged.
We see from this figure that from the intensity of $2\times 10^{6}~\text{V/cm}$ and above, the lifetime is constant, whatever the laser frequency, so that the predominant color in the figure is the dark blue color, which almost indicates, according to the bar-legend, the value of the lifetime when the laser is absent. This means that in this interval there is no effect of the laser on the lifetime, regardless of the laser frequency used. This is due to the fact that the number of exchanged photons $-5\leq s\leq +5$, in this case, is sufficient to achieve the sum-rule. Under the value $2\times 10^{6}~\text{V/cm}$, we note that the lifetime changes according to the intensity and the frequency of the laser field, because the number of photons exchanged cannot reach the sum rule below this value.
So far, we have examined the laser effect on the decay rate and lifetime of the pion, as these two quantities are important in studying decays. Another quantity, not less important, is the branching ratio that was introduced and defined at the end of the outline of the theory. In what follows, we will see the effect of the laser on the branching ratio of both electron and muon.
\begin{figure}[hbtp]
\centering
\includegraphics[scale=0.55]{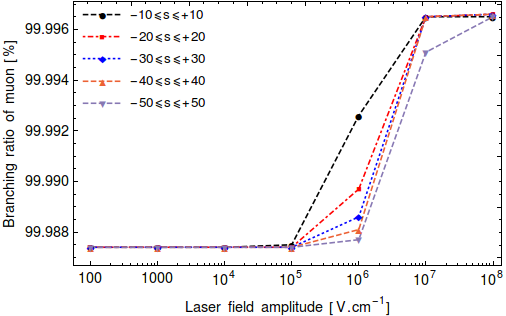}
\caption{The behavior of the branching ratio (\ref{brmuon}) of the muonic decay channel as a function of the laser field amplitude strength for different numbers of photons exchanged. The frequency of laser field is $\hbar\omega=1.17~\text{eV}$.}\label{muonratio}
\end{figure}
\begin{figure}[hbtp]
\centering
\includegraphics[scale=0.55]{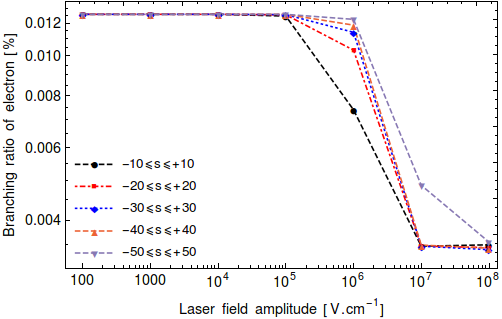}
\caption{The behavior of the branching ratio (\ref{brelec}) of the electronic decay channel as a function of the laser field amplitude for different numbers of photons exchanged. The frequency of laser field is $\hbar\omega=1.17~\text{eV}$. }\label{elecratio}
\end{figure}
Figures \ref{muonratio} and \ref{elecratio} depict the behavior of $\text{Br}(\pi^{-}\rightarrow \mu^{-}\bar{\nu}_{\mu})$ (\ref{brmuon}) and $\text{Br}(\pi^{-}\rightarrow e^{-}\bar{\nu}_{e})$ (\ref{brelec}) versus the laser field amplitude for different numbers of photons exchanged. We notice from Figs.~\ref{muonratio} and \ref{elecratio} that the branching ratio for all numbers of exchanged photons remains constant in the interval of intensities between $10^{2}$ and $10^{5}~\text{V/cm}$ where all the curves are identical. Outside this interval, we note that the branching ratio for the muon increases, and the branching ratio for the electron decreases until they both reach a saturated value at which they  stagnate and all curves meet. According to these results, it becomes clear to us that the branching ratio, unlike the lifetime, is affected by the laser field even if the number of photons exchanged is equal to the cutoff numbers ($s=-50$ and $s=+50$ in this case). Comparing Figs.~\ref{muonratio} and \ref{elecratio}, we note, as required, the branching ratios for both the muon and electron are complementary, as their sum is equal to $100\%$. Another very important remark to make here is that the probability of decay to the muon is much greater than the probability of decay to the electron so that the latter becomes almost non-existent. This fact also applies to the absence of the laser field  where the electronic decay channel must be strongly suppressed by
considering the helicities of the participating particles (see \cite{J} for more details). Therefore, the same thing occurs in  the presence of the laser field, since, at high intensities the branching ratio for the muon increases approximately to $99.996\%$ and that for the electron decreases to $0.003\%$.
\begin{figure}[hbtp]
\centering
\includegraphics[scale=0.55]{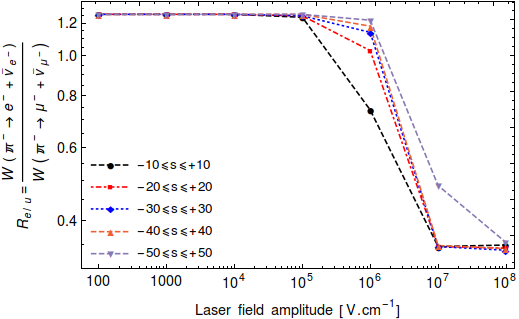}
\caption{The behavior of the branching ratio $R_{e/ \mu}$ (\ref{ratioexp}) (in units of $10^{-4}$) as a function of the laser field amplitude for different numbers of photons exchanged. The frequency of laser field is $\hbar\omega=1.17~\text{eV}$.}\label{ratioelecmuon}
\end{figure}
Figure \ref{ratioelecmuon} illustrates the behavior of $R_{e/ \mu}$ (\ref{ratioexp}) as a function of the laser field amplitude  for different numbers of photons exchanged. It appears that this figure is exactly similar to the Fig.~\ref{elecratio}, meaning that $R_{e/ \mu}\approx\text{Br}(\pi^{-}\rightarrow e^{-}\bar{\nu}_{e})$ and thus $W(\pi^{-}\rightarrow e^{-}\bar{\nu}_{e})\ll W(\pi^{-}\rightarrow \mu^{-}\bar{\nu}_{\mu})$. This is in full agreement with everything that has been discussed previously.
\begin{figure}[hptb]
\subfloat[]{\label{r1}\includegraphics[height=5.5cm,width=.5\linewidth]{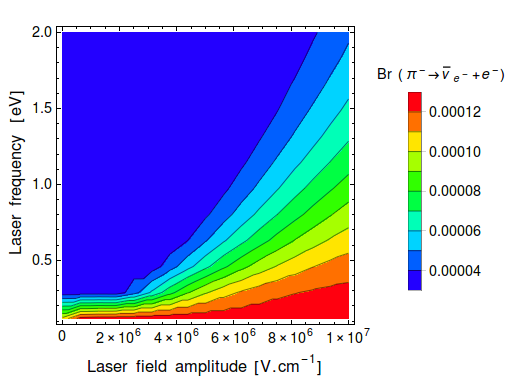}}\hspace*{.5cm}
\subfloat[]{\label{r2}\includegraphics[height=5.5cm,width=.5\linewidth]{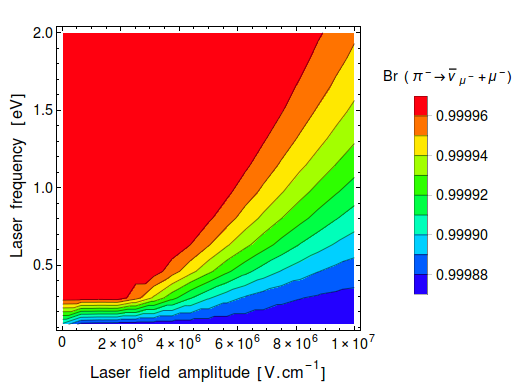}}
\caption{The behavior of the branching ratios as a function of the laser field amplitude $\mathcal{E}_{0}$ and laser frequency $\omega$ for an exchange of $-20\leq s\leq +20$ photons. The laser field amplitude varies from $10^{4}$ to $10^{7}~\text{V/cm}$ whereas the frequency varies from $0.117$ to $2~\text{eV}$. (a) $\text{Br}(\pi^{-}\rightarrow e^{-}\bar{\nu}_{e})$, (b) $\text{Br}(\pi^{-}\rightarrow \mu^{-}\bar{\nu}_{\mu})$.}
\label{ratios}
\end{figure}
Figure \ref{ratios} displays a contour plot of the two branching ratios over the $(\mathcal{E}_{0},\omega)$ plane for $-20\leq s\leq +20$ photons exchanged. With a small number of photons exchanged (for example: $-5\leq s\leq +5$), the branching ratio presents some spots (with minimum or maximum value of the branching ratio at fixed laser field amplitude and frequency) in some locations which represent oscillations inherent to the presence of the ordinary Bessel functions $J_s(z)$ (this situation is not presented here). By increasing the number of photons exchanged to $-20\leq s\leq +20$ and as shown in Fig.~\ref{ratios}, the behavior of the branching ratio has changed and there is no longer an appearance of these spots. Apparently, these two figures are divided into many distinct regions by contours and with different colors accompanied by a value on the bar-legend.  Examining these two figures, we notice that they are superposable, but the same region varies its value from one branching ratio to the other, so that if we combine the two different values for the same region, one obtains the value $1$. Therefore, the two branching ratios are complementary as we have seen before.
\section{Conclusions}
We have performed the analytical calculation for the negatively charged pion decay in the presence of a circularly polarized laser field. Summing up the results, it can be concluded that the pion lifetime (then the decay rate) can be affected by the laser field as long as the number of photons exchanged between the decaying system and the laser field is not sufficient to achieve the well-known sum-rule. Once the sum-rule is checked for a determined number of photons, the influence of the laser on the pion lifetime becomes zero. We have explained the modification of the pion lifetime in the presence of a laser field by considering the well-known quantum Zeno effect. Concerning the branching ratio, we have shown that the laser field increases the probability of decay toward the muonic channel more than the electronic one.

\end{document}